\documentclass[twocolumn,pra,showpacs,aps]{revtex4}

\usepackage{amsmath}
\usepackage{amssymb}
\usepackage{latexsym}
\usepackage{graphicx}
\usepackage[dvips]{color}

\begin{document}
\title{Initial state dependence of the quench dynamics in integrable quantum systems. \\II. Thermal states}
\author{Kai He}
\affiliation{Department of Physics, Georgetown University, Washington, DC 20057, USA}
\author{Marcos Rigol}
\affiliation{Department of Physics, Georgetown University, Washington, DC 20057, USA}

\begin{abstract}
We study properties of isolated integrable quantum systems after a sudden quench starting from thermal
states. We show that, even if the system is initially in thermal equilibrium at finite temperature,
the diagonal entropy after a quench remains a fraction of the entropy in the generalized ensembles
introduced to describe integrable systems after relaxation. The latter is also, in general, different
from the entropy in thermal equilibrium. Furthermore, we examine the difference between the distribution
of conserved quantities in the thermal and generalized ensembles after a quench and show that they
are also, in general, different from each other. This explains why these systems fail to thermalize.
A finite size scaling analysis is presented for each quantity, which allows us making predictions for
thermodynamically large lattice sizes.
\end{abstract}
\maketitle

\section{introduction}\label{sec:introduction}

The recent renewed interest in the theoretical study of the relaxation dynamics and description after
relaxation of isolated many-body quantum systems is largely motivated by extraordinary advances in
experiments with ultracold atomic gases, which provide highly tunable set ups with very high degree
of isolation from the environment
\cite{GreinerNature02,KinoshitaNature06,HofferberthNature07,will_best_10,trotzky_chen_12}. Among many
remarkable experimental results, one can mention the collapse and revival of matter-wave coherence
after the sudden quench from a superfluid to a Mott insulator \cite{GreinerNature02,will_best_10},
the finding that one-dimensional (1D) Bose gases can relax to nonthermal equilibrium states
\cite{KinoshitaNature06}, and the almost perfect agreement between the nonequilibrium dynamics
observed in experiments with 1D lattice bosons and numerically exact simulations of the unitary
dynamics using the time-dependent renormalization group approach \cite{trotzky_chen_12}.

Motivated by the findings in Ref.~\cite{KinoshitaNature06}, intensive theoretical efforts have been
devoted to understanding the description after relaxation of isolated integrable systems following a
sudden change in a Hamiltonian parameter (sudden quench). In quenches between integrable systems, in
which the initial state is an eigenstate of an integrable Hamiltonian (usually the ground state),
many studies have shown that observables do not relax to the thermal predictions
\cite{rigol_dunjko_07_27,cazalilla_06,
rigol_muramatsu_06_26,calabrese_cardy_07a,rigol_dunjko_08_34,eckstein_kollar_08,kollar_eckstein_08,
iucci_cazalilla_09,rossini_silva_09,rossini_susuki_10,fioretto_mussardo_10,iucci_cazalilla_10,mossel_caux_10,
cassidy_clark_11_56,calabrese_essler_11,rigol_fitzpatrick_61,cazalilla_iucci_12,imambekov_preprint_12}.
Instead, they relax to the predictions of generalized Gibbs ensembles (GGEs), which take into account
the presence of nontrivial sets of conserved quantities
\cite{rigol_dunjko_07_27,cazalilla_06,rigol_muramatsu_06_26,calabrese_cardy_07a,rigol_dunjko_08_34,
eckstein_kollar_08,kollar_eckstein_08,iucci_cazalilla_09,rossini_silva_09,rossini_susuki_10,
fioretto_mussardo_10,iucci_cazalilla_10,mossel_caux_10,cassidy_clark_11_56,calabrese_essler_11,
rigol_fitzpatrick_61,cazalilla_iucci_12,imambekov_preprint_12,cramer_dawson_08,barthel_schollwock_08}.
The GGE maximizes the entropy while taking into account the constraints imposed by those conserved
quantities \cite{rigol_dunjko_07_27,jaynes_57a,jaynes_57b}.

Microscopically, the GGE has  recently been shown to work for 1D lattice hard-core bosons
\cite{cassidy_clark_11_56} because of a generalization of the eigenstate thermalization hypothesis
\cite{DeutschPRA91,SrednickiPRE94,rigol_dunjko_07_27}. The idea is that the eigenstates of the final
Hamiltonian that overlap with the initial state have very similar expectation values of the conserved
quantities (properly accounted for in the GGE) and very similar expectation values of few-body
observables, i.e., ETH is valid within that restricted set of eigenstates \cite{cassidy_clark_11_56}.
Furthermore, for quenches very close to integrable points, there is a separation of time scales in
which the system usually relaxes to a prethermalized state before thermalizing
\cite{berges_borsanyi_04,moeckel_kehrein_08, eckstein_kollar_09,moeckel_kehrein_09}. Observables in
that prethermalized state have been shown to be described by a GGE \cite{KollarPRB11}. Interestingly,
for quenches from eigenstates of nonintegrable systems to an integrable point, it has been argued
that thermalization can occur because the initial state provides an unbiased sampling of the
eigenstates in the final Hamiltonian, very much as the thermal ensembles do
\cite{rigol_srednicki_12_70}.

As mentioned before, most works that have directly probed the nonequilibrium dynamics of isolated
integrable systems have focused on initial states that are eigenstates (usually the ground state) of
the Hamiltonian before the quench. This opens questions as to how general the conclusions obtained in
those studies are for more generic finite-temperature systems. Thermal initial states were considered
for quenches in the quantum Ising model in Ref.~\cite{deng_ortiz_11}. There, it was shown that for
sufficiently high initial temperatures nearly thermal distributions occur for the conserved
quantities after a sudden quench. Since away from the critical point the quantum Ising model is
gapped, ``sufficiently high'' in this case means higher than the values of the relevant gaps.

In this work, we study what happens for quenches starting from equilibrium finite-temperature states
in 1D lattice hard-core bosons (the $XY$ model). Within this model, a local chemical potential (a
site-dependent $z$ magnetic field) can be used to generate gaps in the spectrum. We focus on the
resulting entropy and distribution of conserved quantities after a sudden quench and compare them to
the predictions of thermal ensembles and the GGE relevant to the final system in equilibrium.
Understanding the outcome of the relaxation dynamics for initial states at finite temperature is
particularly important to address current ultracold gases experiments, where quenches are performed
starting with the gas at some effective finite temperature.

In the previous and closely related work, we studied the same quantities and quenches similar to the
ones considered here but starting from the ground state of the initial Hamiltonian
\cite{rigol_fitzpatrick_61}. We showed that, if the initial state was the ground state of a
half-filled system in a period 2 superlattice (an insulating state), the distribution of conserved
quantities after the superlattice was turned off approached that in thermal equilibrium upon
increasing the superlattice strength. (An understanding of this phenomenon in terms of bipartite
entanglement was recently provided in Ref.~\cite{cazalilla_preprint_12}.) At the same time, the
entropy of the GGE approached that of the thermal ensemble (grand-canonical ensemble; GE). However,
contrary to what happens in nonintegrable systems, the difference between the GGE and the thermal
ensemble predictions for the entropy and the conserved quantities did not vanish in the thermodynamic
limit for any finite strength of the superlattice. For all other quenches considered in that work,
the predictions of the GGE and the thermal ensembles were quantitatively and qualitatively different.

Here we show that following a sudden quench and upon increasing the temperature of the initial state,
similarly to what was found in Ref.~\cite{rigol_fitzpatrick_61} when increasing the strength of the
initial superlattice potential and consistent with the findings in Ref.~\cite{deng_ortiz_11} for the
quantum Ising model, the distribution of conserved quantities and the entropy in the GGE approach the
thermal predictions. However, this should not be confused with thermalization, as, for any given
initial finite temperature and finite Hamiltonian parameters before and after the quench, the
differences between the two ensembles remain finite in the thermodynamic limit. Hence, our results
show that for quenches between integrable systems, there is no fundamental difference between
starting with an eigenstate of the Hamiltonian and starting with a state in thermal equilibrium.
Thermalization does not generally occur in either case.

The exposition is organized as follows. In Sec.~\ref{sec:model}, we introduce the model of interest,
and the definition of the statistical ensembles and the observables studied. In
Sec.~\ref{sec:weight}, we discuss the behavior of the weights of the eigenstates of the final
Hamiltonian in various ensembles. The energy density distributions are also studied in that section.
Subsequently, in Sec.~\ref{sec:entropy}, we study various entropies and perform a series of scaling
analysis to assess finite-size effects as well as the influence of the control parameters in the
properties of the system after the quench. A similar study, but for the behavior of the conserved
quantities, is presented in Sec.~\ref{sec:conserved quantity}. Finally, our results are summarized in
Sec.\,\ref{sec:conclusion}.

\section{Model, ensembles, and observables}\label{sec:model}
We focus on the nonequilibrium properties of 1D lattice hard-core bosons following a sudden quench.
This is an integrable model with Hamiltonian
\begin{equation}\label{eq:hamiltonian}
    \hat{H} = -t\sum^{L-1}_{i=1} (\hat{b}^\dag_i \hat{b}^{}_{i+1}+\textrm{H.c.})+ A \sum^L_{i=1}
(-1)^i\,
\hat{n}_i,
\end{equation}
where $t$ is the hopping parameter and $A$ is the strength of a local alternating (superlattice)
potential. We consider lattices with $L$ sites and open boundary conditions. The hard-core boson
creation (annihilation) operators are denoted  $\hat{b}^\dag_i(\hat{b}_i)$, and the number operator
$\hat{n}_i=\hat{b}^\dag_i\hat{b}_i$. In addition to the bosonic commutation relations $[\hat{b}_i,
\hat{b}^\dag _j]=\delta_{ij}$, there applies a constraint that suppresses multiply occupancy of the
lattice sites in all physical states, i.e., $\hat{b}^{\dag 2}_i=\hat{b}^2_i=0$. We note that, in the
following, $\hbar=1,\,k_B=1$, and the hopping energy is set to $t=1$ (our unit of energy).

To be solved exactly, this model can first be  mapped onto the spin-1/2 $XY$ Hamiltonian through the
Holstein-Primakoff transformation \cite{Holstein40} and, subsequently, via the Jordan-Wigner
transformation \cite{Jordan28}, to noninteracting fermions. Instead of diagonalizing the full
many-body Hamiltonian, one can then write each many-body eigenstate in the form of a fermionic Slater
determinant with appropriate modifications following the mapping rules. These Slater determinants are
constructed as products of single-particle eigenstates of the noninteracting fermionic Hamiltonian.
Utilizing properties of Slater determinants, one can compute exactly all one-particle
\cite{rigol_muramatsu_04_8,rigol_muramatsu_05_13} and two-particle \cite{HePRA11, HePRA12}
observables in the eigenstates, at finite temperature in the GE \cite{rigol_05_19}, and out of
equilibrium \cite{rigol_muramatsu_05_16}.

Before the quench, our system is assumed to have $N$ particles and to be in contact with a thermal
reservoir at finite temperature $T$ in equilibrium. At time $\tau=0$, we disconnect the system from
the reservoir and change the strength of the alternating potential from its initial value $A_I$ to
its final value $A_F$ ($\hat{H}_I\rightarrow\hat{H}_F$). The initial thermal state, which is a mixed
state, can be expressed in the basis of the many-body eigenstates $|\Psi^I_\alpha\rangle$ of the
initial Hamiltonian $\hat{H}_I$, $\hat{H}_I|\Psi^I_\alpha\rangle=E^I_\alpha|\Psi^I_\alpha\rangle$.
The many-body density matrix can be written as
\begin{equation}\label{eq:init dens matx}
    \hat{\rho}_I = \frac{1}{Z_I} \sum_\alpha e^{-E^I_\alpha/{T}} |\Psi^I_\alpha\rangle \langle
\Psi^I_\alpha|,
\end{equation}
where $Z_I=\sum_\alpha e^{-E^I_\alpha/{ T}}$. The time evolution of the density matrix is given by
\begin{equation}\label{eq:time dens matx}
  \hat{\rho}(\tau) = \frac{1}{Z_I} \sum_\alpha e^{-E^I_\alpha/{ T}} |\Psi^I_\alpha(\tau)\rangle
\langle \Psi^I_\alpha(\tau)|,
\end{equation}
in which the time-evolving many-body eigenstates of the initial Hamiltonian
$|\Psi^I_\alpha(\tau)\rangle$ can be written in terms of the many-body eigenstates of the final
Hamiltonian, $\hat{H}_F|\Psi^F_\beta\rangle=E^F_\beta|\Psi^F_\beta\rangle$:
\begin{eqnarray}\label{eq:time eigenstate}
    |\Psi^I_\alpha(\tau)\rangle &=& e^{-i \hat{H}_F\tau}|\Psi^I_\alpha\rangle, \nonumber\\
    &=& \sum_\beta |\Psi^F_\beta\rangle \;e^{-i E^F_\beta \tau}
\langle\Psi^F_\beta|\Psi^I_\alpha\rangle.
\end{eqnarray}
Following Eq.~\eqref{eq:time eigenstate}, the infinite time average of Eq.\ \eqref{eq:time dens matx}
can be written in the form of a diagonal density matrix, which corresponds to the so-called diagonal
ensemble (DE) \cite{rigol_dunjko_07_27},
\begin{equation}
      \overline{\hat{\rho}}= \lim_{{\tau'}\to\infty} \frac1{\tau'}\int_0^{\tau'} d\tau
\hat{\rho}(\tau)
= \hat{\rho}_{\textrm{DE}}\equiv \sum_\beta W_\beta
|\Psi^F_\beta\rangle \langle\Psi^F_\beta|, \label{eq:dens matx DE}
\end{equation}
where we have assumed that degeneracies, if present, are irrelevant, and have defined
\begin{equation}
    W_\beta =\frac{1}{Z_I} \sum_\alpha e^{-E^I_\alpha/{
T}}|\langle\Psi^F_\beta|\Psi^I_\alpha\rangle|^2.\label{eq:weight DE}
\end{equation}
Here, $W_\beta$ corresponds to the weight of state $|\Psi^F_\beta\rangle$ in the DE. As we show
below, $W_\beta$ is strongly dependent of the initial state temperature and the quench protocol. In
Eq.~\eqref{eq:weight DE}, the overlaps between eigenstates of the initial and those of the final
Hamiltonian $\langle\Psi^F_\beta|\Psi^I_\alpha\rangle$ are evaluated numerically as the determinant
of the product of two matrices, with each matrix composed of the matrix elements of the Slater
determinant representing the eigenstates \cite{rigol_fitzpatrick_61}. Note that the dimension of the
Hilbert space for our model is $d=\binom{L}{N}$, and the computation time for the entire set of DE
weights is proportional to $d^2$, i.e., it is exponential in the system size.

The expectation value of any observable $\hat{O}$ after relaxation, if relaxation takes place, is
equal to the infinite-time average $\overline{\hat{O}}$ and can be calculated as
$\text{Tr}[\hat{\rho}_{\textrm{DE}}\, \hat{O}]$ \cite{cassidy_clark_11_56}. Following the definition
of the von Neumann entropy $S=-\textrm{Tr}[\hat\rho\ln\hat\rho]$, one can also calculate the entropy
of the DE as
\begin{equation}\label{eq:entropy DE}
    S_{\textrm{DE}}=-\sum_\beta W_\beta \ln(W_\beta).
\end{equation}
This entropy has  recently been shown to satisfy the required properties of a thermodynamic entropy,
namely, it increases when a system in equilibrium is taken out of equilibrium, it satisfies the
fundamental thermodynamic relation \cite{Polkovnikov11}, and it is additive and equal (up to
subextensive corrections) to the entropy of thermal ensembles used to describe generic systems after
relaxation \cite{SantosPRL11}.

Here, the DE is compared with various statistical ensembles. If the number of particles $N$ in the
system is kept fixed and the energy is allowed to fluctuate about a mean value determined by the
initial state $E_I=\textrm{Tr}[\hat\rho_I\hat H_{F}]$, the relevant ensemble to compare with is the
canonical ensemble (CE). The density matrix in this case has the form
\begin{equation}\label{eq:dens matx CE}
    \hat\rho_{\textrm{CE}}=\frac{1}{Z_{\textrm{CE}}} \sum_\beta e^{-E^F_\beta/{
T_{\textrm{CE}}}}|\Psi^F_\beta\rangle \langle\Psi^F_\beta|,
\end{equation}
where $Z_{\textrm{CE}}=\sum_\beta e^{-E^F_\beta/{T_{\textrm{CE}}}}$ and $T_{\textrm{CE}}$ is taken so
that $\textrm{Tr}[\hat\rho_{\textrm{CE}}\hat H_{F}]=E_I$. Similarly, for a system in which not only
the energy but also the number of particles is allowed to fluctuate, the relevant ensemble is the GE,
for which the density matrix has the form
\begin{equation}\label{eq:dens matx GE}
    \hat\rho_{\textrm{GE}}=\frac{1}{Z_{\textrm{GE}}} \sum_\beta e^{-(E^F_\beta-\mu N^F_\beta)/{
T_{\textrm{GE}}}}|\Psi^F_\beta\rangle \langle\Psi^F_\beta|,
\end{equation}
where $Z_{\textrm{GE}}=\sum_\beta e^{-(E^F_\beta-\mu N^F_\beta)/{T_{\textrm{GE}}}}$ and $N^F_\beta$
is given by $\hat N |\Psi^F_\beta\rangle=N^F_\beta|\Psi^F_\beta\rangle$ ($\hat N$ is the operator for
the total number of particles). Note that, in Eq.~\eqref{eq:dens matx GE}, the sum runs over the
entire set of many-body eigenstates of $\hat H_F$ with all possible particle numbers.
$T_{\textrm{GE}}$ and $\mu$ need to be chosen so that $\textrm{Tr}[\hat\rho_{\textrm{GE}}\hat
H_F]=E_I$ and $\textrm{Tr}[\hat\rho_{\textrm{GE}} \hat N]=N$. The entropies corresponding to each of
those ensembles can be directly calculated using von Neumann's definition and yield
\begin{align}
     &S_{\textrm{CE}}=\ln Z_{\textrm{CE}}+\frac{E_I}{ T_{\textrm{CE}}}, \label{eq:entropy CE}\\
     &S_{\textrm{GE}}=\ln Z_{\textrm{GE}}+\frac{E_I-\mu N}{ T_{\textrm{GE}}}. \label{eq:entropy
GE}
\end{align}
These two entropies agree with each other in the thermodynamic limit, for any given value of $N$ and
$E$, and the same is true for the predictions of both ensembles (and the microcanonical ensemble) for
additive observables. In the GE, because of the mapping between hard-core bosons and fermions,
$Z_{\textrm{GE}}$ can be calculated in terms of the single-particle eigenenergies of the final
Hamiltonian $\epsilon_n$, and $T_{\textrm{GE}}$ and $\mu$ \cite{rigol_05_19}:
\begin{equation}
 Z_{\textrm{GE}}=\prod_{n=1}^L \left[ 1+e^{-(\epsilon_n-\mu)/T_{\textrm{GE}}} \right].
\end{equation}

In isolated integrable systems, however, the presence of nontrivial sets of constants of motion
constrain the dynamics and lead to expectation values of observables after relaxation that are
different from the predictions of thermal ensembles. The GGE is then the natural choice to describe
such systems \cite{rigol_dunjko_07_27}. The many-body density matrix of the GGE can be written as
\begin{equation}\label{eq:dens GGE}
    \hat \rho_{\textrm{GGE}}=\frac{1}{Z_{\textrm{GGE}}} e^{-\sum_n \lambda_n \hat I_n},
\end{equation}
where $Z_{\textrm{GGE}}=\textrm{Tr}\big[e^{- \sum_n \lambda_n \hat I_n}\big]$ and, for hard-core
bosons, $\{\hat I_n\}$ are nothing but the projection operators to the single-particle eigenstates of
$\hat H_F$. (The fact that these quantities are conserved is evident if we write the final
Hamiltonian as $\hat{H}_F=\sum_n \epsilon_n \hat I_n$.) $\{\lambda_n\}$ are Lagrange multipliers,
which need to be selected to meet the initial conditions $\textrm{Tr}[\hat\rho_I \hat
I_n]=\textrm{Tr}[\hat\rho_{\textrm{GGE}} \hat I_n]$. For hard-core bosons, they can be computed as
\cite{rigol_dunjko_07_27}
 \begin{equation}
\lambda_n=\ln \left[\frac{1-\textrm{Tr}[\hat\rho_I \hat I_n]}{\textrm{Tr}[\hat\rho_I \hat
I_n]}\right]
\label{eq:lagrangem}
 \end{equation}
The GGE entropy can then be written as
\begin{equation}\label{eq:entropy GGE}
    S_{\textrm{GGE}}=\ln Z_{\textrm{GGE}}+\sum_n \lambda_n \textrm{Tr}[\hat\rho_I \hat
I_n].
\end{equation}

In this work, we focus on two types of quenches; the first one is the turning-off of the superlattice
potential $A_I\neq0$, $A_F=0$, and the second one is the reverse of the first one, namely, the
turning-on of the superlattice potential $A_I=0$, $A_F\neq0$. In all cases, we start from a
finite-temperature mixed state.

\section{weights in the ensembles}\label{sec:weight}

We are first interested in learning how the weights $W_\beta$ are distributed among different
eigenstates, how they depend on the initial temperature and quenching protocol, and how they compare
to the weights of the eigenstates in the CE. Clearly, if $W_\beta$ values approach the CE weights
$Z^{-1}_{\textrm{CE}} e^{-E^F_\beta/{ T_{\textrm{CE}}}}$, thermalization will take place.

\begin{figure}[!b]
    \includegraphics[width=0.50\textwidth, angle=-90]{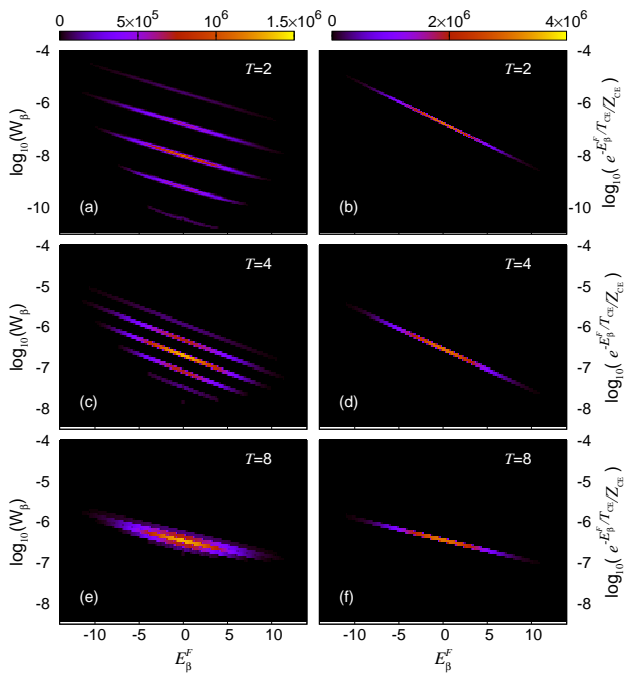}
\vspace{-0.1cm}
\caption{(Color online) Density plot of the coarse-grained weights of energy eigenstates in the
DE $W_\beta$ (a, c, e) and in the CE
$Z^{-1}_{\textrm{CE}} e^{-E^F_\beta/{ T_{\textrm{CE}}}}$ (b, d, f). Results presented
are for lattices with $L=24$, $N=12$ (half filling) and for quenches from $A_I=4$ to $A_F=0$.
The values of the initial temperature are: $T=2$ (a, b), $T=4$ (c, d),
and $T=8$ (e, f). The color scale indicates the number of states per unit area.
}\label{fig:weight 1}
\end{figure}

In Fig.~\ref{fig:weight 1}, we show the coarse-grained weights of the eigenstates of the final
Hamiltonian in the DE [Figs.~\ref{fig:weight 1}(a), \ref{fig:weight 1}(c), and \ref{fig:weight 1}(e)]
and the CE [Figs.~\ref{fig:weight 1}(b), \ref{fig:weight 1}(d), and \ref{fig:weight 1}(f)]. The
results are obtained for quenches $A_I=4$ to $A_F=0$, and for three different initial temperatures.
One can see there that, at the lowest temperature $(T=2)$, the distribution of weights in the DE
[Fig.~\ref{fig:weight 1}(a)] is very different from that in the CE [Fig.~\ref{fig:weight 1}(b)].
$W_\beta$ exhibits a banded structure that is in stark contrast with the simple exponential decay
seen in the CE. Remarkably, as the temperature in the initial state increases, as shown in
Figs.~\ref{fig:weight 1}(c) and \ref{fig:weight 1}(d), the distance between the bands seen in the DE
decreases and the slopes of the bands approach that of the CE. For even higher initial temperatures,
higher than the gaps in the initial state ($\Delta\propto A_I$) [Figs.~\ref{fig:weight 1}(e) and
\ref{fig:weight 1}(f)], the bands in the DE merge and the weights very closely resemble the canonical
weights, i.e., the expectation value of observables after relaxation in such quenches will be nearly
thermal. It is not difficult to understand this because, in the $T\rightarrow\infty$ limit, the
initial thermal state will have a completely flat (featureless) distribution of conserved quantities.
Such a distribution of conserved quantities coincides with the one at infinite temperature in thermal
equilibrium after the quench. Hence, the initial state will essentially provide an unbiased sampling
of the eigenstates that make the main contribution to the CE.

\begin{figure}[!b]
    \includegraphics[width=0.51\textwidth, angle=-90]{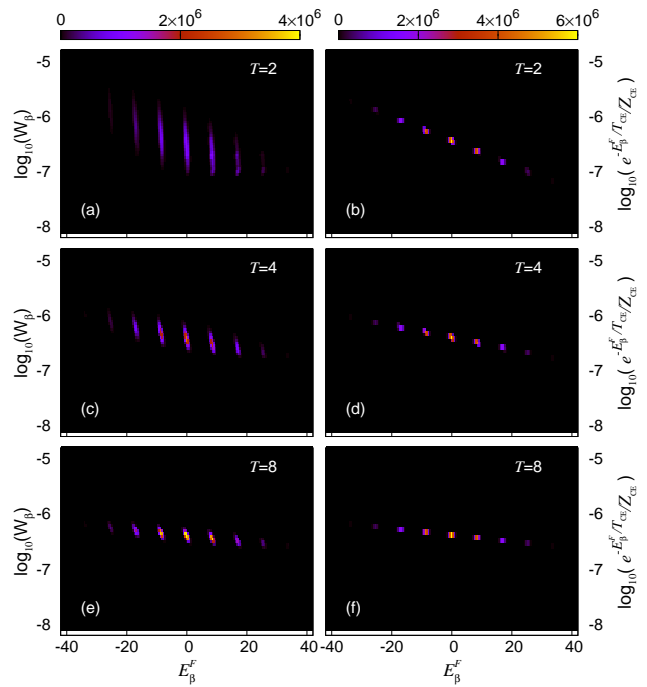}
\vspace{-0.1cm}
\caption{(Color online) Same as Fig.~\ref{fig:weight 1} but for quenches from $A_I=0$
to $A_F=4$.}\label{fig:weight 2}
\end{figure}

Results for the reverse quench to the one in Fig.~\ref{fig:weight 1}, namely, a quench from $A_I=0$
to $A_F=4$, are depicted in Fig.~\ref{fig:weight 2}. Due to the energy gaps present in the many-body
spectrum of the final Hamiltonian (generated by the presence of the superlattice potential), the
distribution of weights in both the DE and the CE exhibit bands separated by gaps $\Delta\propto
A_F$. Once again, within each band, the weights in the DE and the CE are very different at low
temperatures ($T=2$ in Fig.~\ref{fig:weight 2}). However, it is also apparent in Fig.~\ref{fig:weight
2} that these weights become similar to each other as the temperature of the initial state increases.

\begin{figure}[!b]
    \includegraphics[width=0.48\textwidth]{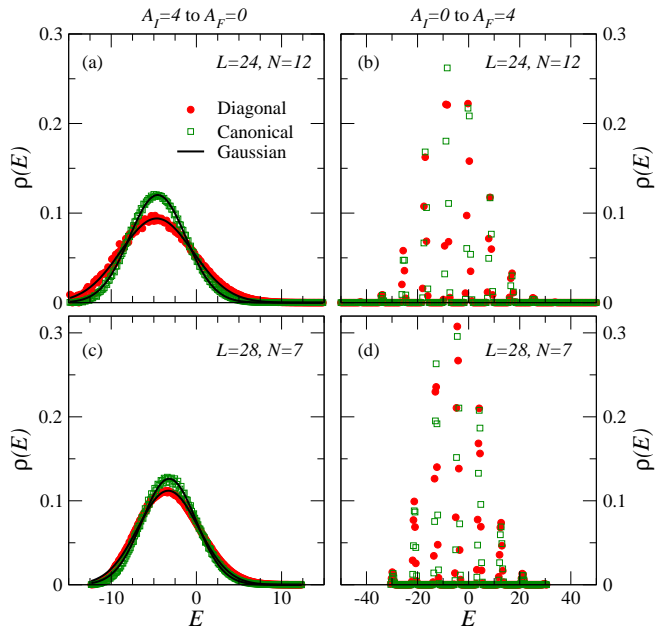}
\vspace{-0.3cm}
\caption{(Color online) Energy density $\rho(E)$ for two quenches at finite $T=2$. (a, c) A quench
from $A_I=4$ to $A_F=0$; (b, d) a quench from $A_I=0$ to $A_F=4$. (a, b) A
half-filled system ($N=L/2$); (c, d) a quarter-filled one ($N=L/4$).
In each panel, we report the results for $\rho(E)$ in the DE and the CE.
In (a) and (c), regardless of the filling, it is apparent that the energy distributions
in both ensembles have a Gaussian shape. Hence, we also report results for a Gaussian fit
$\rho(E)=(\sqrt{2\pi}\, \delta E)^{-1} e^{-(E-E_I)^2/(2\, \delta E^2)}$ to each curve.}\label{fig:energy
dens}
\end{figure}

To better quantify the contribution of each part of the energy spectrum to the different ensembles,
we have extracted out of Figs.\ \ref{fig:weight 1} and \ref{fig:weight 2} the energy density
$\rho(E)$. This quantity is proportional to the sum of the weights in a given energy window times the
number of states in that window. In our plots, the results for $\rho(E)$ are properly normalized such
that the integral of $\rho(E)$ over the full spectrum is unity. Results for that quantity, and $T=2$,
are presented in Fig.~\ref{fig:energy dens}. It is remarkable to see that, for quenches from $A_I=4$
to $A_F=0$ [Fig.\ \ref{fig:energy dens}(a) and (b)], the energy density in the DE exhibits a Gaussian
shape independently of the filling of the system.

In Ref.~\cite{rigol_fitzpatrick_61}, results for quenches starting from the ground state showed that
only for half-filled systems and large values of $A_I$ does a Gaussian shape develop in $\rho(E)$ in
the DE. Here we find that, at finite and not too low initial temperatures, the energy densities as
well as other observables are qualitatively similar for different filling factors. This is in strong
contrast with what happens for quenches starting from the ground state. Such a contrast is expected,
as the ground states for different fillings may be qualitatively different, e.g., insulating at
half-filling and superfluid away from half-filling, but those differences are washed out with
increasing temperature. In Fig.~\ref{fig:energy dens}, note that the Gaussian like shape observed in
the DE is slightly wider that in the CE. However, the width of both Gaussians is expected to vanish
in the thermodynamic limit.

The results for quenches from $A_I=0$ to $A_F=4$ (depicted in the Figs.~\ref{fig:energy dens}) also
develop a Gaussian shape in both ensembles. However, this is less evident because of the presence of
gaps in the spectrum. For these quenches, we also find that the results for the various observables
studied here are qualitatively similar for different fillings when the temperatures are not too low.
Because of this, in the remainder of the paper we focus on the half-filled case.

\section{entropies}\label{sec:entropy}

We showed in the previous section that, for initial states at finite temperature, the energy
distribution after quenches within integrable systems takes a Gaussian-like form. This is interesting
because, for quenches starting from pure states that are eigenstates of the initial Hamiltonian,
Gaussian-like energy distributions are only generically observed in nonintegrable systems
\cite{SantosPRL11}. Since the energy density is calculated through a coarse graining of the weights,
in the following, we calculate the entropy associated with the DE and compare it to the entropy in
the GE and the GGE. This allows us to better quantify the number of states contributing to each
ensemble and to assess whether thermalization can take place in the system.

\begin{figure}[!b]
    \includegraphics[width=0.48\textwidth]{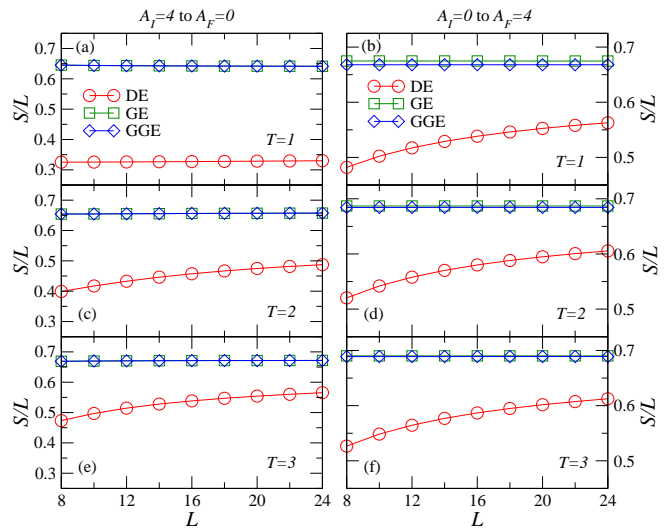}
\vspace{-0.3cm}
\caption{(Color online) Entropy per site as a function of system size for various ensembles and
initial temperatures. (a, c, e) Quenches from $A_I=4$ to $A_F=0$; (b, d, f) Quenches from
$A_I=0$ to $A_F=4$. The systems are at half-filling and
the initial temperatures are $T=1$ (a, b), $T=2$ (c, d), and $T=3$ (e, f).
}\label{fig:entropy}
\end{figure}

In Fig.\ \ref{fig:entropy}, we show the entropy per site with increasing $L$ for the DE, GE, and GGE,
and for quenches starting from different initial temperatures. There are several important
conclusions that can be reached from those results. (i) For all entropies and quenches, $S/L$
saturates with increasing system size, as expected since $S$ is an additive quantity. (ii) For all
quenches at finite temperature, $S_{\textrm{DE}}$ is a fraction of $S_{\textrm{GE}}$ and
$S_{\textrm{GGE}}$, and our finite-size scaling analysis indicates that this will be the case in the
thermodynamic limit, i.e., exponentially fewer states are involved in the DE compared to the other
two. This is consistent with the findings in Refs.~\cite{cassidy_clark_11_56, SantosPRL11,
rigol_fitzpatrick_61} for quenches starting from pure states. (iii) Also, for all quenches,
$S_{\textrm{GE}}$ and $S_{\textrm{GGE}}$ are very close to each other. They can actually be seen to
approach each other further as the temperature in the initial state increases.

The lack of agreement between the entropy of the DE and that of the GGE does not mean that the GGE
will fail to describe observables after relaxation. As shown in Ref.~\cite{cassidy_clark_11_56}, the
great majority of eigenstates in the DE and generalized ensembles have very similar expectation
values of few-body observables, i.e., independently of the number of states in each ensemble, the
expectation value of the observables will coincide. What is of more interest in the remainder of the
paper is how the GGE compares with the GE. An agreement between the two means that the integrable
system would actually thermalize.

\begin{figure}[tb]
    \includegraphics[width=0.48\textwidth]{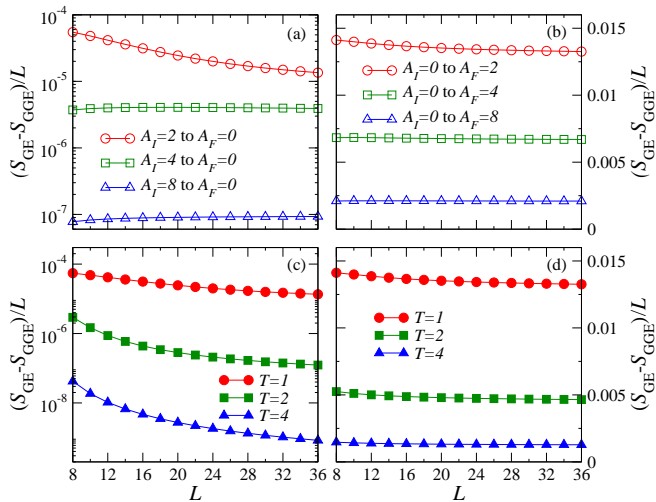}
\vspace{-0.3cm}
\caption{(Color online) Difference between $S_{\textrm{GGE}}$ and $S_{\textrm{GE}}$ per site as a
function of $L$ for quenches $A_I\neq0\rightarrow A_F=0$ (a, c) and for quenches
$A_I=0\rightarrow A_F\neq0$ (b, d). (a, b) Results obtained for
a fixed initial temperature $T=1$ but different values of $A_I$ (a) and $A_F$ (b). (c, d)
Results obtained for fixed $A_I=2$ (c) and $A_F=2$ (d) but different values of $T$.
All systems are at half-filling.
}\label{fig:entropy difference}
\end{figure}

In Figs.~\ref{fig:entropy difference}(a) and \ref{fig:entropy difference}(b), we plot a finite-size
scaling of $(S_{\textrm{GE}}-S_{\textrm{GGE}})/L$ for systems with the same initial temperature but
quenched between different values of $A_I$ and $A_F$. In all cases, the difference is found to
saturate to a finite value with increasing system size. Consistent with the findings in
Ref.~\cite{rigol_fitzpatrick_61} for quenches starting from the ground state, we do find that as the
value of $A_I$ increases, for  quenches $A_I\neq0\rightarrow A_F=0$ [Fig.~\ref{fig:entropy
difference}(a)], or as the value of $A_F$ increases, for  quenches $A_I=0\rightarrow A_F\neq0$
[Fig.~\ref{fig:entropy difference}(b)], the difference between the two entropies decreases.

In Figs.~\ref{fig:entropy difference}(c) and \ref{fig:entropy difference}(d), we plot a finite-size
scaling of $(S_{\textrm{GE}}-S_{\textrm{GGE}})/L$ for systems with fixed values of $A_I$ and $A_F$
but for different initial temperatures. Similar to the results in Figs.~\ref{fig:entropy
difference}(a) and \ref{fig:entropy difference}(b), we find that the difference saturates
to a finite value with increasing system size. However, that value decreases as the temperature in
the initial state increases, and can become negligibly small.

From the results depicted in Fig.\ \ref{fig:entropy difference} we arrive at the conclusion that,
even though the difference between the entropy of the GGE and that of the GE becomes negligibly small
if some control parameters (the values of $A_I$ or $A_F$, and of $T$) are changed, it never vanishes
in the thermodynamic limit as long as those control parameters are kept finite and fixed. This means
that the GE and the GGE do not become equivalent in the thermodynamic limit, and thermalization will
not generally occur at a finite temperature.

\begin{figure}[tb]
\begin{center}
    \includegraphics[width=0.45\textwidth]{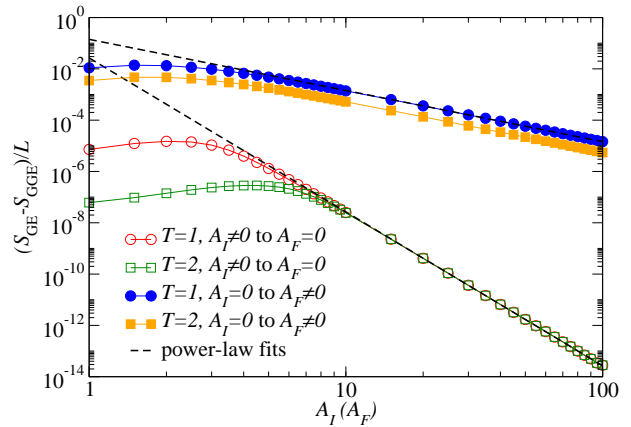}
\end{center}
\vspace{-0.4cm}
\caption{(Color online) $(S_{\textrm{GE}}-S_{\textrm{GGE}})/L$ as a function of $A_I$ for
quenches $A_I\neq0\rightarrow A_F=0$ (open symbols) and $A_F$ for $A_I=0\rightarrow A_F\neq0$
(filled symbols), with the initial temperature fixed. For each quench, results for two
temperatures are presented. The dashed lines signal the power-law decay observed for
large values of $A_I(A_F)$: $A^{-6}_I$ for the quench $A_I\neq0\rightarrow A_F=0$, and $A^{-2}_F$
for the quench $A_I=0\rightarrow A_F\neq0$. The systems are at half-filling with $L=32$ in all
cases.
}\label{fig:scaling}
\end{figure}

In Fig.~\ref{fig:scaling}, we show how the difference between the entropy per site in the GE and the
GGE changes with increasing $A_I$ or $A_F$ while the initial temperature is kept fixed. For both
quenches, $A_I\neq0\rightarrow A_F=0$ and $A_I=0\rightarrow A_F\neq0$,
$(S_{\textrm{GE}}-S_{\textrm{GGE}})/L$ exhibits a power-law decay in the regime of large $A_I(A_F)$.
When the systems are quenched by switching off the superlattice potential, the results for different
$T$ values are on top of each other once $A_I$ becomes sufficiently large, and they decay at $\sim
1/A_I^6$. The exponent of this power law is the same found in quenches from the ground in
Ref.~\cite{rigol_fitzpatrick_61}. This is because the half-filled system in the presence of a
superlattice exhibits a gap $\Delta=2A_I$ between the ground state and the first excited state, so as
long as the temperature is much lower than the gap, the initial system is essentially in its ground
state. (The ground state in the limit $A_I\rightarrow\infty$ is a trivial Fock state, namely, the
product of empty and occupied single-site states.) For the quench $A_I=0\rightarrow A_F\neq0$, one
can see in Fig.~\ref{fig:scaling} that the results for different temperatures do not coincide with
each other. However, they do exhibit the same power-law decay,
$(S_{\textrm{GE}}-S_{\textrm{GGE}})/L\sim 1/A_F^2$, independently of the temperature of the initial
state. The vanishing of the difference between the entropy of the GGE and the GE as $A_I(A_F)$
increases indicates that the ensembles become equivalent to each other and observables will exhibit a
thermal-like behavior after relaxation.
\begin{figure}[tb]
\begin{center}
    \includegraphics[width=0.45\textwidth]{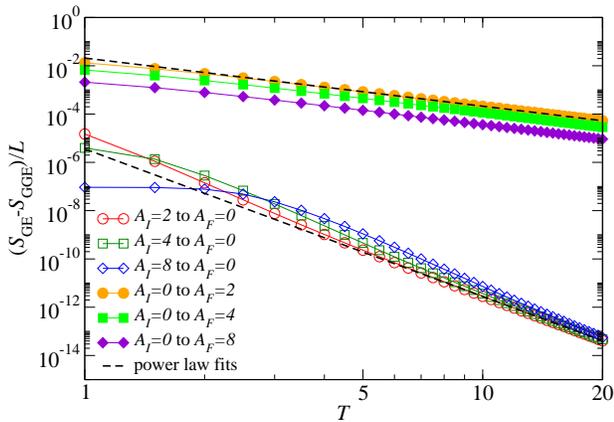}
\end{center}
\vspace{-0.4cm}
\caption{(Color online) $(S_{\textrm{GE}}-S_{\textrm{GGE}})/L$ as a function of $T$ for
quenches $A_I\neq0\rightarrow A_F=0$ (open symbols) and $A_I=0\rightarrow A_F\neq0$
(filled symbols), for fixed values of $A_I$ or $A_F$. For each kind of quench, results for three
values of nonzero $A_I(A_F)$ are presented. Dashed lines are power-law fits to quenches
$A_I=2\rightarrow A_F=0$ [$(S_{\textrm{GE}}-S_{\textrm{GGE}})/L\sim 1/T^{6.11}$] and
$A_I=0\rightarrow A_F=2$ [$(S_{\textrm{GE}}-S_{\textrm{GGE}})/L\sim 1/T^{1.99}$], in the region
$T>10$. The systems are at half filling with $L=32$ in all cases.
}\label{fig:scaling 2}
\end{figure}

In Fig.~\ref{fig:scaling 2}, we study how $(S_{\textrm{GE}}-S_{\textrm{GGE}})/L$ behaves as a
function of the temperature of the initial state for different quenches $A_I\neq0\rightarrow A_F=0$
and $A_I=0\rightarrow A_F\neq0$. There one can see that the difference between the entropy per site
in the GE and that in the GGE decreases with increasing $T$. The decay is power-law-like for high
values of $T$, and is faster for quenches $A_I\neq0\rightarrow A_F=0$ than for quenches
$A_I=0\rightarrow A_F\neq0$. We have fitted our results for $T>10$ to a power law in the quenches
$A_I=2\rightarrow A_F=0$ obtaining $(S_{\textrm{GE}}-S_{\textrm{GGE}})/L\sim 1/T^{6.11}$ and in the
quenches $A_I=0\rightarrow A_F=2$ obtaining $(S_{\textrm{GE}}-S_{\textrm{GGE}})/L\sim 1/T^{1.99}$.

\section{conserved quantities}\label{sec:conserved quantity}

In this section, we study the expectation values of the conserved quantities $\hat{I}_n$ in the GGE
and in the GE as one changes system parameters, similarly to what was done in Sec.\ \ref{sec:entropy}
for the entropy. The conserved quantities $\{\hat{I}_n\}$ considered here are the set of $L$
projection operators to the single-particle eigenstates of the final noninteracting fermionic
Hamiltonian to which hard-core bosons can be mapped. By construction, the expectation values of the
conserved quantities in the GGE are identical to those in the initial state, i.e., $\langle \hat
I_n\rangle_{\textrm{GGE}}=\textrm{Tr}[\rho_I \hat I_n]$. In the GE, they can be computed
straightforwardly because the occupations of the single-particle energy levels follow the Fermi-Dirac
distribution $\langle\hat I_n\rangle_{\textrm{GE}}=1/\big[1+e^{(\epsilon_n-\mu)/{
T_{\textrm{GE}}}}\big]$, where $\epsilon_n$ are the single-particle eigenenergies.

\begin{figure}[!b]
    \includegraphics[width=0.48\textwidth]{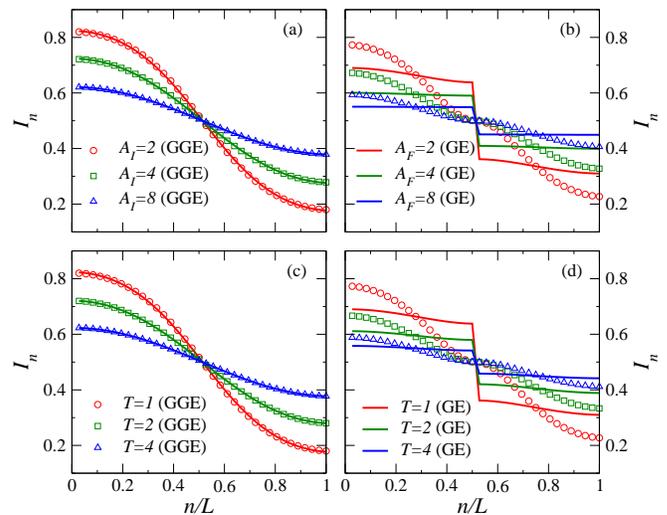}
    \vspace{-0.4cm}
\caption{(Color online)
Expectation value of the conserved quantities $I_n=\langle\hat{I}_n\rangle$ corresponding to the
$n$th lowest energy eigenstate in the single-particle spectrum. Results are presented for
quenches $A_I\neq0\rightarrow A_F=0$ (a, c), and for quenches
$A_I=0\rightarrow A_F\neq0$ (b, d). (a, b) Results for systems with the same
initial temperature $T=1$ but different values of $A_I$ (a) and $A_F$ (b); (c, d)
results for systems with the same $A_I=2$ (c) and $A_F=2$ (d) but different
initial temperatures. In all panels, open symbols depict GGE results, and solid lines
depict GE results. All systems are at half-filling with $L=36$.}
\label{fig:conserved quantities}
\end{figure}

In Fig.\ \ref{fig:conserved quantities}, we plot the expectation values of the conserved quantities
in the GGE and the GE for different combinations of $A_I(A_F)$ and $T$, for both types of quenches
studied in the previous sections. Figures \ref{fig:conserved quantities}(a) and \ref{fig:conserved
quantities}(b) depict results for quenches with the same initial temperature but different values of
$A_I(A_F)$. When the half-filled system is quenched from $A_I\neq0$ to $A_F=0$, $\langle \hat
I_n\rangle_{\textrm{GGE}}$ and $\langle \hat I_n\rangle_{\textrm{GE}}$ are almost indistinguishable
from each other and they are smooth functions of $n$ [Fig.\ \ref{fig:conserved quantities}(a)]. The
picture for quenches $A_I=0\rightarrow A_F\neq0$ is very different [Fig.\ \ref{fig:conserved
quantities}(b)]. In this case, the expectation values of the conserved quantities in the GE exhibits
a discontinuity at $n/L=0.5$, which is the result of the gap opened by the superlattice potential. On
the other hand, $\langle \hat I_n\rangle_{\textrm{GGE}}$ is a smooth function of $n$. However, the
presence of the band gap does have an effect on $\langle \hat I_n\rangle_{\textrm{GGE}}$, as it tends
to flatten out its values in the vicinity of the gap position (an effect that becomes more evident as
$A_F$ increases). This effect is only seen at finite temperatures, as in the ground state $\langle
\hat I_n\rangle_{\textrm{GGE}}$ is the same between quenches from $A_I\neq0$ and quenches to
$A_F\neq0$ when $A_I$ in the former equals $A_F$ in the latter \cite{rigol_fitzpatrick_61}.
Note that, for both kinds of quenches, the expectation values of all conserved quantities approach
a constant (1/2) value as $A_I$ or $A_F$ increases.

\begin{figure}[t]
    \includegraphics[width=0.48\textwidth]{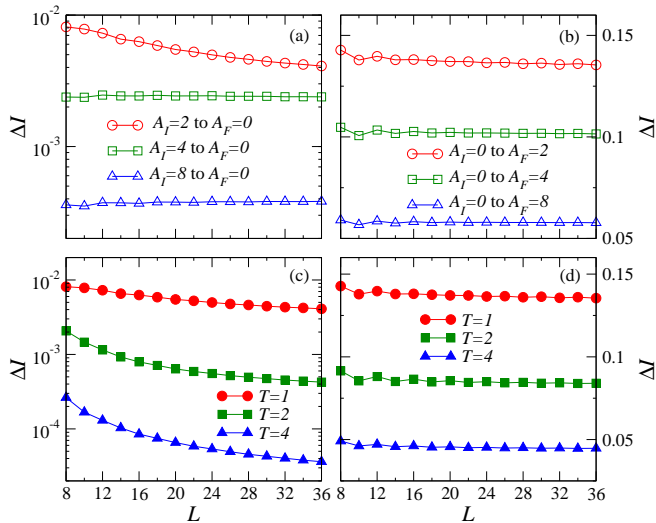}
\vspace{-0.4cm}
\caption{(Color online)
Integrated difference $\Delta I$ as a function of $L$ for quenches $A_I\neq0\rightarrow A_F=0$ (a, c)
and for quenches $A_I=0\rightarrow A_F\neq0$ (b, d). Parameters are the same as
in Fig.\ \ref{fig:entropy difference}. All systems are at half-filling.}\label{fig:integrated
difference}
\end{figure}

In Fig.\ \ref{fig:conserved quantities}, we show the results for quenches $A_I=2\rightarrow A_F=0$
[Fig.\ \ref{fig:conserved quantities}(c)] and $A_I=0\rightarrow A_F=2$ [Fig.\ \ref{fig:conserved
quantities}(d)] in systems with different values of the initial temperature. The overall picture is
similar to that in the top panels when $A_I(A_F)$ is changed. As the initial temperature increases,
the expectation values of all conserved quantities approach a constant (1/2) value, while the
specific features of each kind of quench are still visible. Namely, for quenches $A_I=2\rightarrow
A_F=0$, the GGE and thermal predictions are very close to each other and, for quenches
$A_I=0\rightarrow A_F=2$, a discontinuity is always seen in the GE prediction. Such a discontinuity
in the latter quenches is absent in the distribution of conserved quantities in the initial state.

In order to be more quantitative, and for comparison with the results obtained for the entropy, we
calculate the relative integrated difference between the GGE and the GE predictions for the conserved
quantities
\begin{equation} \label{eq:int diff}
    \Delta I=\frac{\sum_n |\langle \hat I_n\rangle_{\textrm{GGE}}-\langle \hat
I_n\rangle_{\textrm{GE}}|}{\sum_n \langle \hat I_n\rangle_{\textrm{GGE}}},
\end{equation}
Scaling results for $\Delta I$ vs $L$ are presented in Fig.\ \ref{fig:integrated difference}. They
are qualitatively similar to those for the entropy differences in Fig.\ \ref{fig:entropy difference}.
In all quenches studied, the differences  between the GGE and the GE predictions for the entropies
and the conserved quantities are seen to saturate to a finite value with increasing system size.
Also, the differences between the GGE and the GE predictions for the conserved quantities decrease as
$A_I$ increases (for quenches $A_I\neq0\rightarrow A_F=0$) and as $A_F$ increases (for quenches
$A_I=0\rightarrow A_F\neq0$), as well as when $T$ increases.

\begin{figure}[tb]
    \includegraphics[width=0.45\textwidth]{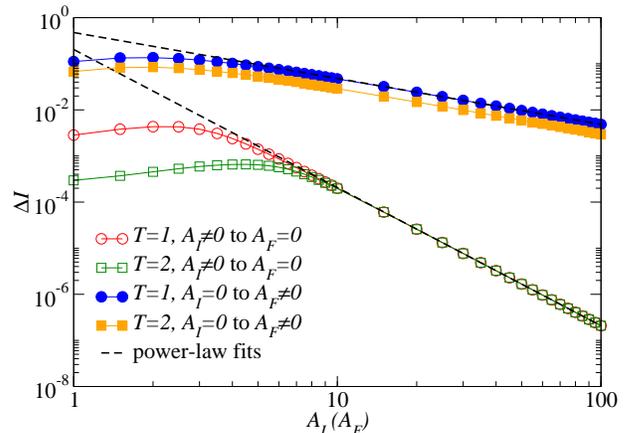}
\vspace{-0.4cm}
\caption{(Color online)
$\Delta I$ as a function of $A_I$, for quenches $A_I\neq0\rightarrow A_F=0$ (open symbols),
and of $A_F$, for quenches $A_I=0\rightarrow A_F\neq0$ (filled symbols), for two  values
of $T$. Dashed lines depict power-law fits to the large $A_I$ ($\Delta I\sim 1/{A^3_I}$)
and $A_F$ ($\Delta I\sim 1/{A_I}$) results. Systems are at half-filling with $L=32$.}
\label{fig:cq scaling}
\end{figure}

The dependence of $\Delta I$ on $A_I$($A_F$), for a fixed system size and for two different
temperatures, is depicted in Fig.~\ref{fig:cq scaling}. For both kinds of quenches, one can see that
$\Delta I$ vanishes as a power law in the regime of large $A_I(A_F)$: for the quench from $A_I\neq0$
to $A_F=0$, $\Delta I$ decreases as $\Delta I\sim 1/{A^3_I}$, while for the opposite quench it
decreases as $\Delta I \sim 1/{A_F}$. One should note that, as expected from the discussion in
Sec.~\ref{sec:entropy}, the behavior seen for $A_I\gg T$ is independent of $T$ and identical to that
in the ground state. The exponents of the power laws are also independent of $T$ for both quenches
and are found to be one-half those for $(S_{\textrm{GE}}-S_{\textrm{GGE}})/L$ vs $A_I(A_F)$.

\begin{figure}[tb]
    \includegraphics[width=0.45\textwidth]{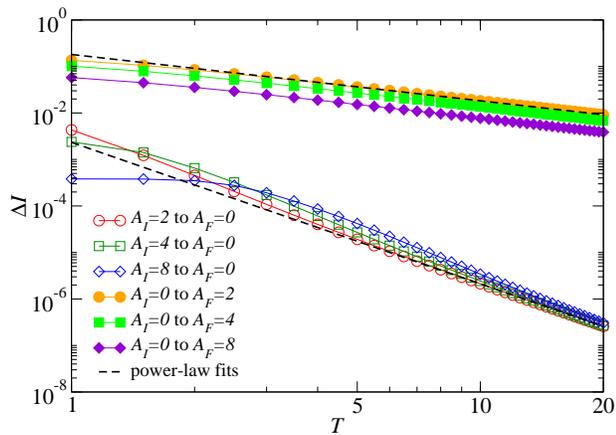}
\vspace{-0.4cm}
\caption{(Color online)
$\Delta I$ as a function of $T$ for quenches $A_I\neq0\rightarrow A_F=0$ (open symbols) and
quenches $A_I=0\rightarrow A_F\neq0$ (filled symbols). Dashed lines are power-law fits in the region
$T>10$. We obtain $\Delta I \sim 1/T^{3.04}$ for quenches $A_I=2\rightarrow A_F=0$ and
$\Delta I\sim 1/T^{1.00}$ for quenches $A_I=0\rightarrow A_F=2$, respectively.
All systems are at half-filling with $L=32$.}
\label{fig:cq scaling 2}
\end{figure}

$\Delta I$ as a function of $T$ is shown in Fig.\ \ref{fig:cq scaling 2}, for three quenches
$A_I\neq0\rightarrow A_F=0$ and for three quenches $A_I=0\rightarrow A_F\neq0$. The behavior of
$\Delta I$ is once again similar to that observed for $(S_{\textrm{GE}}-S_{\textrm{GGE}})/L$ vs $T$
in the same quenches; $\Delta I$ vanishes as a power law at very high temperatures. Power-law fits in
the region $T>10$ yield $\Delta I\sim 1/T^{3.04}$ for quenches $A_I=2\rightarrow A_F=0$ and $\Delta
I\sim 1/T^{1.00}$ for quenches $A_I=0\rightarrow A_F=2$. Once again, the exponents are found to be
one-half those for $(S_{\textrm{GE}}-S_{\textrm{GGE}})/L$ vs $T$ in Sec.\ \ref{sec:entropy}.

The preceding analysis clearly shows that only when the difference between the conserved quantities
in the initial state (in the GGE) and those in the thermal ensembles becomes negligible do the
entropies and the ensembles themselves become equivalent. An understanding of this, as well as of the
relation between the exponents of the power-law decays seen for
$(S_{\textrm{GE}}-S_{\textrm{GGE}})/L$ and $\Delta I$, can be gained if one realizes that the entropy
in the GGE and the GE can be written in terms of the occupations of the single-particle eigenstates
as
\begin{equation}
 S=-\sum_{n=1}^L [(1-I_n) \ln(1-I_n)+I_n\ln I_n],
\end{equation}
where ${I_n}=\langle \hat I_n\rangle_{\textrm{GGE}}$ for the GGE and ${I_n}=\langle \hat
I_n\rangle_{\textrm{GE}}$ for the GE. Once the occupations ${I_n}$ in both ensembles are very close
to each other, they are also very close to 1/2 (see Fig.~\ref{fig:conserved quantities}). One can
write $\langle \hat I_n\rangle_{\textrm{GE}}-1/2=\varepsilon^\text{GE}_n$ and $\langle \hat
I_n\rangle_{\textrm{GGE}}-1/2=\varepsilon^\text{GGE}_n$ (with $|\varepsilon_n|\ll 1$); it is then
straightforward to show that
\begin{align}
 &S_\textrm{GE}-S_\textrm{GGE}
=\sum_{n=1}^L\Big\{2[(\varepsilon^\text{GGE}
_n)^2-(\varepsilon^\text{GE\,}_n)^2] \nonumber\\
&+\frac{4}{3}[(\varepsilon^\text{GGE}_n)^4-(\varepsilon^\text{GE\,}_n)^4] 
+O[(\varepsilon^\text{GGE}_n)^6-(\varepsilon^\text{GE\,}_n)^6]\Big\} ,
\end{align}
which provides an excellent description of the results for $(S_\textrm{GE}-S_\textrm{GGE})/L\ll 1$ in
Sec.~\ref{sec:entropy} and decays with a power-law exponent that is twice that for $\Delta I$.

\section{conclusions} \label{sec:conclusion}
In this paper, we have studied properties of 1D lattice hard-core bosons ($XY$ chain) after quenches
that start from initial thermal states. In all cases considered, the quenches were generated by
sudden changes in a superlattice potential (a local space-dependent magnetic field in the spin
language). We have shown that, in these integrable systems after a quench starting from a
finite-temperature state,  the coarse-grained energy densities exhibit a Gaussian shape. However, the
distributions of the weights are still qualitatively different between the quenched state and thermal
states. This dissimilarity leads to an extensive difference among the entropies in the isolated
system after the quench (the diagonal entropy), the GGE, and the thermal ensembles. On the other
hand, as one (or both) of the two control parameters explored here (the initial temperature $T$ and
the strength of the superlattice $A_I$ or $A_F$) are tuned to infinity, all ensembles become
equivalent. The approach between the entropy of the GE and that of the GGE under such tuning was
shown to be power law in the control parameter, independent of the quench protocol selected and of
the values of the other parameters that were kept fixed. It is important to emphasize that, when all
parameters in the quenches were kept fixed and finite and the system size was extrapolated to
infinity, the differences between the GGE and the GE results were seen to saturate at a finite value,
i.e., these two ensembles do not become equivalent.

We have also shown that such differences have their origin in the disagreement between the conserved
quantities after the quench, which are determined by the initial state, and the distribution of
conserved quantities in thermal ensembles. By tuning the control parameters mentioned above to
infinity, we have seen that the distribution of the conserved quantities in the quenched state
approaches that in thermal equilibrium, which explains why the generalized and thermal ensembles
approach each other under those conditions. However, we should stress that, in our particular systems
of interest and quench protocols followed, thermalization only occurs when the control parameters are
tuned to lead to completely flat distributions of conserved quantities, which is what happens when
the initial temperature is infinite or when $A_I(A_F)\rightarrow\infty$.

\begin{acknowledgments}
This work was supported by the U.S. Office of Naval Research. We thank Juan Carrasquilla for useful
discussions.
\end{acknowledgments}


\end{document}